\documentclass[english]{article}
\usepackage[T1]{fontenc}
\usepackage[utf8]{inputenc}
\usepackage{geometry}
\usepackage{color}
\usepackage{babel}
\usepackage{float}
\usepackage{textcomp}
\usepackage{amstext}
\usepackage{graphicx}
\usepackage{setspace}
\usepackage[unicode=true,pdfusetitle,
 bookmarks=true,bookmarksnumbered=false,bookmarksopen=false,
 breaklinks=false,pdfborder={0 0 1},backref=false,colorlinks=false]
 {hyperref}
\begin{document}
\begin{doublespace}
\begin{center}
\textbf{\LARGE{}Michael Milken: The Junk Dealer}{\LARGE\par}
\par\end{center}

\begin{center}
\textbf{Ravi Kashyap}
\par\end{center}

\begin{center}
\textbf{SolBridge International School of Business / City University
of Hong Kong}
\par\end{center}

\begin{center}
July-04-2019
\par\end{center}

\begin{center}
\par\end{center}

\begin{center}
Financial Economics 
\par\end{center}

\begin{center}
\tableofcontents{}
\par\end{center}
\end{doublespace}
\begin{doublespace}

\section{Abstract}
\end{doublespace}

\begin{doublespace}
We take a closer look at the life and legacy of Micheal Milken. We
discuss why Michael Milken, also know as the Junk Bond King, was not
just any other King or run-of-the-mill Junk Dealer, but ”The Junk
Dealer”. We find parallels between the three parts to any magic act
and what Micheal Milken did, showing that his accomplishments were
nothing short of a miracle. His compensation at that time captures
to a certain extent the magnitude of the changes he brought about,
the eco-system he created for businesses to flourish, the impact he
had on the wider economy and also on the future growth and development
of American Industry. We emphasize two of his contributions to the
financial industry that have grown in importance over the years. One
was the impetus given to the Private Equity industry and the use of
LBOs. The second was the realization that thorough research was the
key to success, financial and otherwise. Perhaps an unintended consequence
of the growth in junk bonds and tailored financing was the growth
of Silicon valley and technology powerhouses in the California bay
area. Investors witnessed that there was a possibility for significant
returns and that financial success could be had due to the risk mitigation
that Milken demonstrated by investing in portfolios of so called high
risk and low profile companies. We point out the current trend in
many regions of the world, which is the birth of financial and technology
firms and we suggest that finding innovative ways of financing could
be the key to the sustained growth of these eco-systems.
\end{doublespace}
\begin{doublespace}

\section{\label{sec:The-Price-Junk}The Price of Junk }
\end{doublespace}

\begin{figure}[H]
\includegraphics[width=17cm]{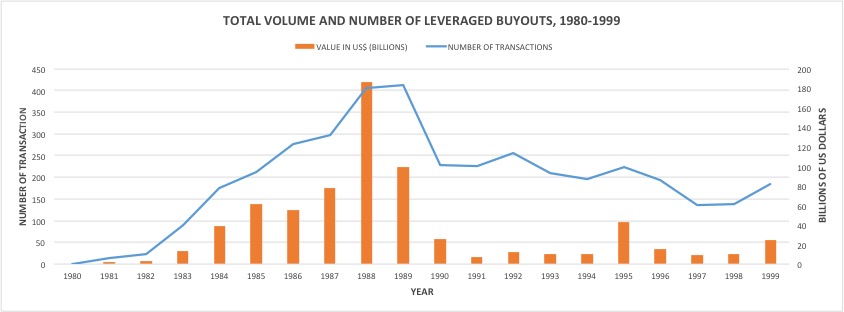}

\caption{\label{fig:The-Market-for-Junk}The Market for Junk}
\end{figure}

\begin{doublespace}
The life and legacy of Micheal Milken, his meteoric rise and subsequent
fall from fame, can be summarized by the words, ``Someone’s Garbage
is Someone Else’s Treasure''. His compensation (salary and bonus)
from 1983 to 1984 jumped from \$43 million to \$123 million reaching
a peak of \$500 million USD in 1987 (Bruck 1989; Sobel 2000; End-note
\ref{enu:The-following-sources}). During this time, the Junk Bond
Market grew from almost non-existent to more than \$125 billion USD
(Figure \ref{fig:The-Market-for-Junk}; Altunbaş, Gadanecz \& Kara
2006a, 2006b; End-note \ref{enu:Figure-Market-Junk}). 

We take a closer look at the larger than life accomplishments of Micheal
Milken and his subsequent rapid decline. We discuss why Michael Milken,
also know as the Junk Bond King, was not just any other King or run-of-the-mill
Junk Dealer, but ”The Junk Dealer”. We find parallels between the
three parts to any magic act and what Micheal Milken did, showing
that his accomplishments were nothing short of a miracle. His compensation
at that time captures to a certain extent the magnitude of the changes
he brought about, the eco-system he created for businesses to flourish,
the impact he had on the wider economy and also on the future growth
and development of American Industry. We emphasize two of his contributions
to the financial industry that have grown in importance over the years.
One was the impetus given to the Private Equity industry and the use
of LBOs (leveraged buyouts: Kaplan \& Stromberg 2009; End-notes \ref{enu:Private-equity-typically},
\ref{enu:A-leveraged-buyout}). The second was the realization that
thorough research was the key to success, financial and otherwise.
Perhaps an unintended consequence of the uptick in junk bonds and
tailored financing was the growth of silicon valley and technology
powerhouses in the California bay area. Investors witnessed that there
was a possibility for significant returns and that financial success
could be had due to the risk mitigation that Milken demonstrated by
investing in portfolios of so called high risk and low profile companies.
We point out the current trend in many regions of the world, which
is the birth of financial and technology firms, and we suggest that
finding innovative ways of financing could be the key to the sustained
growth of these eco systems.

A big influence on how Michael Milken approached investing and in
particular how he assessed the risk of managing portfolios of junk
bonds can be attributed to the book, Corporate Bond Quality and Investor
Experience (Hickman 1958). Looking at his roots, tell us that he probably
had honorable intentions to begin with and lost his way along the
journey. We discuss whether he had crossed a certain point of no return,
where he had to do whatever he had to, so as to sustain his creation.
Some of the topics we discuss are: The Price of Junk (section \ref{sec:The-Price-Junk});
Milken was Milk’em LBOs (Leveraged Buyouts) (section \ref{sec:Milken-was-Milk=002019em});
Rescuing Fallen Angels (section \ref{subsec:Rescuing-Fallen-Angels});
Milken’s Magic: The Pledge; The Turn; The Prestige (section \ref{sec:Milken=002019s-Magic});
The Rise and Fall (section \ref{sec:The-Rise-and}); Milken the Felon
(or Not?) (section \ref{subsec:Milken-the-Felon}); Client Loyalty,
A Capital Crime (?) (section \ref{subsec:Client-Loyalty,-A}); What
would you do? What is your verdict? (section \ref{subsec:What-would-you})

\end{doublespace}
\begin{doublespace}

\subsection{To Yield or Not To Yield}
\end{doublespace}

\begin{doublespace}
Bonds are an attractive way to raise capital since there is no need
to give out firm ownership. The key terminology about bonds which
most of us know, or are sort of familiar with, such as principal,
coupon and maturity of a bond are explained in (Ross, Westerfield
\& Jaffe 2013). The following equation (eq \ref{eq:Bond-Value}) summarizes
the relation between bond prices and discount rates, which are closely
related to interest rates and can also factor in the risk of defaults
on a loan taken by a firm. Clearly we see that there is an inverse
relationship between interest rates and bond prices.
\begin{equation}
\text{Bond Price or Bond Value}=\frac{C}{r}\left[1-\frac{1}{\left(1+r\right)^{T}}\right]+\frac{F}{\left(1+r\right)^{T}}\label{eq:Bond-Value}
\end{equation}
Here, $C$ is the coupon amount; $F$ is the face value; $T$ is the
time to maturity; $r$ is the discount rate for the cash flows that
accrue to the investor in the bond. It is the return the bond holder
would obtain by holding the bond till maturity with all payments made
as scheduled and reinvested at the same rate. It can be viewed as
the required return on the bond and is also known as the yield to
maturity.

Another central aspect of the working of bond markets was the role
played by the credit rating agencies: Fitch, Moody’s and S\&P. These
rating agencies evaluated companies on their financial health and
assigned them letter-grade credit ratings of between C and AAA. Firms
that were rated low (companies in the B range and below were deemed
as having credit risk, that is an inability to meet the payments on
their loans; Kashyap 2019 provides a discussion of the role perceptions
play when we create rankings) by these agencies had significant difficulties
raising capital and bonds issued by these companies were known as
junk bonds. (White 2010) is a recent exploration of how the financial
regulatory structure propelled these three credit rating agencies
to the center of the U.S. bond markets and thereby virtually guaranteed
that when these rating agencies did make mistakes, those mistakes
would have serious consequences for the financial sector.
\end{doublespace}
\begin{doublespace}

\section{\label{sec:Milken-was-Milk=002019em}Milken was Milk’em LBOs (Leveraged
Buyouts) }
\end{doublespace}

\begin{doublespace}
``Finance is an art. Not Yes or No, Right or Wrong. It is an art
form, an understanding of who should be the companies of the future,
and how to structure transactions. It is an art form greatly misunderstood.''
-{}-{}-{}-{}- Michael Milken, August 25, 1986 (End-note \ref{enu:=00201CFinancing-is-an}).
With this fundamental belief, Michael Milken created an eco system
for junk bonds that flourished beyond anyone's expectations. Suffice
it to say he was making the most of the situation (End-note \ref{enu:Ghostface-Killah-=000026})
\end{doublespace}
\begin{doublespace}

\subsection{High School Spirits of A Dynamic Duo}
\end{doublespace}

\begin{doublespace}
Michael Milken was born in Los Angeles, California on the 4th of July
1946, the American Independence Day. He was voted “most spirited”
and “friendliest” class member in the class of 1964, at which time
he was also crowned prom king. Being too short to join the basketball
team, he joined the cheerleading squad, creating controversy at times,
for example, by leading vigorous cheers when his team was ahead 42
to nothing. This should make us reminiscence a bit about basketball,
which was created by guys, who were 5’ 7”, tall and is being played
by guys, who are 7’ 5” tall, with the same basket height. We can infer
from this that he was personable, well liked and had a competitive
nature. 

The rest of what people said may or may not be entirely correct, depending
on when they said it, before or after, Milken’s rise and fall from
fame. “He was extremely bright, with a prodigious memory and a very
competitive spirit. While less athletic than his brother, he exhibited
a leadership charisma very early in life. Some called him a boy wonder”
(Bruck 1989; Sobel 2000; End-note \ref{enu:The-following-sources}).
Lori Ann Hackel, his high school sweetheart, who later became his
wife, was school princess (End-note \ref{enu:Again-School-Princess})
and voted “most likely to succeed” (End-note \ref{enu:Voted-most-likely}). 
\end{doublespace}
\begin{doublespace}

\subsection{A Sweet Deal: Wild Wild West, Starting From the East}
\end{doublespace}

\begin{doublespace}
California has inspired a few hundred Songs (End-note \ref{enu:California-Songs}),
that should tell us something about the kind of motivational effects
it has on people ... Perhaps, a lot? In an op-editorial written at
age 24 in 1970, which Milken submitted to (and was rejected by) the
New York Times, he said, “Unlike other crusaders from Berkeley, I
have chosen Wall Street as my battleground for improving society because
it is here that governments, institutions and industries are financed.”
This suggests that he was starting out with the right intentions about
doing good for society. The alternative viewpoint could be that at
that age, he knew he would go on to create a multi-billion dollar
industry, get involved in numerous scams and regulatory inquiries
and thought it appropriate to send out misleading messages even before
he started. It would be simpler to grant him the benefit of doubt.

He left graduate school and headed to Drexel Firestone, which later
became Drexel Burnham Lambert. Drexel President I.W. Tubby Burnham
offered Milken, “One Dollar for Every Three Dollars” (Sobel 2000)
when he started the job. That is, Milken could keep one dollar for
three dollars profit he made for the company and that deal stayed
the same when Milken was playing with billions. The kind of deal that
would be unheard of these days.

Milken went back to his roots and relocated to Beverly Hills, California.
He initially got an opportunity to be involved in some non-financial
restructuring. As operations analyst, Milken surveyed the various
business units and found a hierarchy topped by sales, then by traders
in the middle, then researchers on the bottom. Buyers for products
existed, but sales staff was inadequate to develop and distribute
new products. Only research could create a broader market. Thus, over
time, he placed an emphasis on research, which was to obtain a proper
understanding of companies, their capital needs and risk factors.
As he rose within the organization, research came to dominate the
organizational order. 
\end{doublespace}
\begin{doublespace}

\subsection{\label{subsec:Rescuing-Fallen-Angels}Rescuing Fallen Angels }
\end{doublespace}

\begin{figure}[H]
\includegraphics[width=17cm]{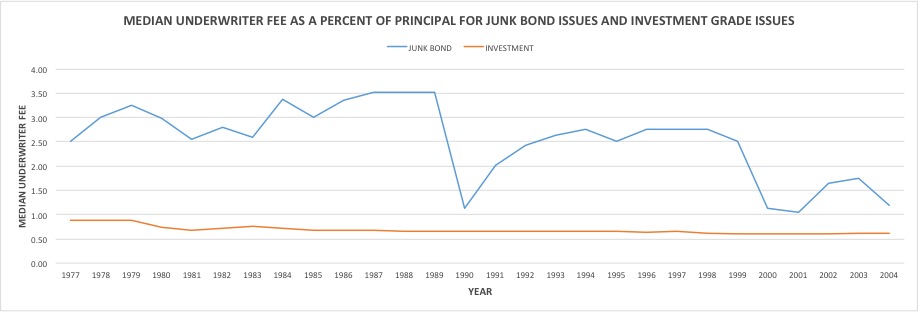}

\caption{\label{fig:Spread-The-Junk}Spread The Junk}
\end{figure}

\begin{figure}[H]
\includegraphics[width=17cm]{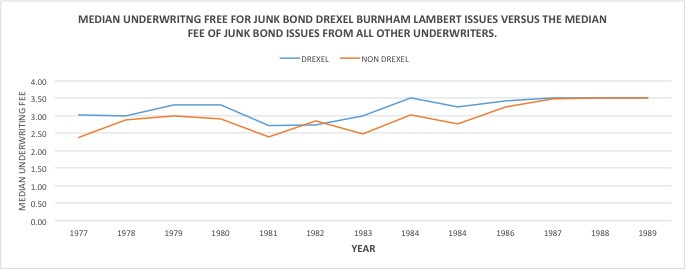}

\caption{\label{fig:Drexel-Underwriting-Fees}Drexel Underwriting Fees}
\end{figure}

\begin{doublespace}
Milken became the primary player in the secondary junk bond market.
He then moved to primary market offerings. The higher risk of default
on low grade bonds was compensated by the higher interest that they
paid (Figures \ref{fig:Spread-The-Junk}, \ref{fig:Drexel-Underwriting-Fees};
Livingston \& Williams 2007; End-note \ref{enu:Figures-Drexel-Fees}).
The market overestimated the cost of companies failing to meet payments
on their obligations. Thus bond prices were too low and the quality
yield spreads correspondingly too high. If fewer bonds actually defaulted
than were expected to default, then a diversified portfolio comprising
such bonds would outperform a portfolio of investment-grade bonds.
The low grade bond market consisted of “fallen angels”: bonds that
had been listed as investment grade but had been downgraded. Milken
took an interest in these companies that had fallen on hard times.
In his eyes, these companies had strong assets to back their balance
sheets and could generate strong earnings. The importance of his team's
approach to understanding the value of companies with extensive research
needs to be emphasized here.

\end{doublespace}
\begin{doublespace}

\subsection{No Easy Credit: A Blessing to Some}
\end{doublespace}

\begin{doublespace}
In 1977, only a handful of non-investment grade companies issued bonds.
Paradoxically, the most innovative companies, those with the highest
returns on capital and fastest rates of growth, had the least access
to capital. Only 6\% of roughly 11,000 corporations were investment
grade. Suffice it to say that banks were not giving any easy credit
to corporations in the wild 1970s. This is ironic since Milken's team
gave him full credit, “Someone like Mike comes along once every five
hundred years” (Bruck 1989; End-note \ref{enu:Again-Credit-Compensation}).
\end{doublespace}
\begin{doublespace}

\subsection{The Tipping Point}
\end{doublespace}

\begin{figure}[H]
\includegraphics[width=12cm]{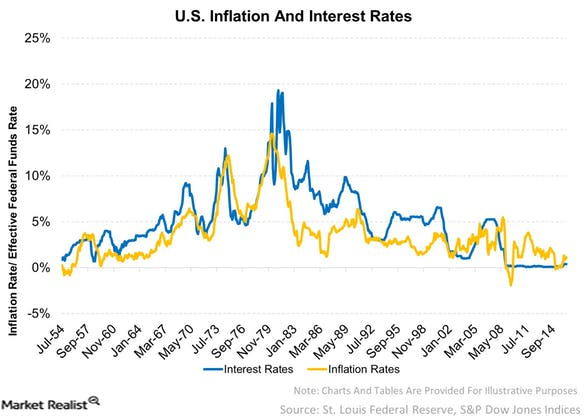}\caption{\label{fig:Interest-Rates-and}Interest Rates and Inflation}
\end{figure}

\begin{doublespace}
From 1977 to 1981 new issues of junk bonds were less than \$1.5 billion
per year. We can identify the tipping point (Gladwell 2006; End-note
\ref{enu:Tipping-Point}) somewhere after this time since from 1982
to 1987 there were junk bonds everywhere (Figure \ref{fig:The-Market-for-Junk}).
(Figure \ref{fig:Interest-Rates-and}) shows that interest rates were
at an all time high around 1980 adding to the issue of limited capital
and high interest payments for companies with low investment ratings.
In 1983, Drexel issued \$1.1 billion in junk bonds for MCI Communications,
another major milestone in the ascent of Milken. By positioning themselves
in the middle of buyers and sellers, Milken and Drexel controlled
the market. Drexel also bought back the bonds they underwrote. This
show of faith was instrumental to building credibility and trust,
key ingredients in financial transactions. 60\% of Drexel profits
in 1985 with underwriting fees of 3\% to 4\% came from junk bonds.

\end{doublespace}
\begin{doublespace}

\section{\label{sec:Milken=002019s-Magic}Milken’s Magic }
\end{doublespace}

\begin{doublespace}
Every great magic trick consists of three parts or acts (End-note
\ref{enu:The-Prestige-is}). The first part is called \textquotedbl The
Pledge\textquotedbl . The magician shows you something ordinary:
a deck of cards, a bird or a man. He shows you this object. Perhaps
he asks you to inspect it to see if it is indeed real, unaltered,
normal. But of course... it probably isn't. The second act is called
\textquotedbl The Turn\textquotedbl . The magician takes the ordinary
something and makes it do something extraordinary. Now you're looking
for the secret... but you won't find it, because of course, you're
not really looking. You don't really want to know. You want to be
fooled. But you wouldn't clap yet because making something disappear
isn't enough; you have to bring it back. That's why every magic trick
has a third act, the hardest part, the part we call \textquotedbl The
Prestige\textquotedbl .'

Let us now try to identify these three parts in whatever Milken did.
If we are able to convince ourselves that these three acts do exist,
then Milken becomes a Magician.
\end{doublespace}
\begin{doublespace}

\subsection{The Pledge of Michael Milken}
\end{doublespace}

\begin{doublespace}
This is the simplest or the most basic and obvious aspect of what
Milken did. He intervened to match buyer and seller and depended on
an intricate network of reciprocal obligation. He would sell the junk
bonds of Corporation X only if it agreed, at least tacitly, to buy
the junk bonds of Corporate Y at an advantageous time. He thought
of himself as a retailer. The duty of a retailer is to talk to a customer
and see what they want. It became clear to Milken that customers needed
products that were not available at that time. The solution was staring
them in the face, they had to create those products. Pledge also means
a promise and Milken was a man of his word, “In my entire career on
Wall Street I have never backed out of a transaction once I have agreed
to stand up to it, no matter how onerous it might be”, (Bruck 1989;
End-note \ref{enu:The-following-sources}).
\end{doublespace}
\begin{doublespace}

\subsection{The Turn of Michael Milken}
\end{doublespace}

\begin{figure}[H]
\includegraphics[width=8cm]{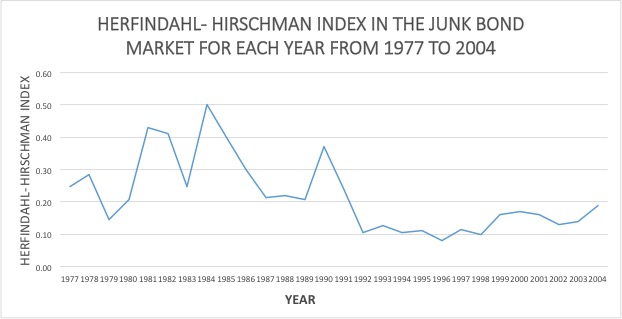}\includegraphics[width=8cm]{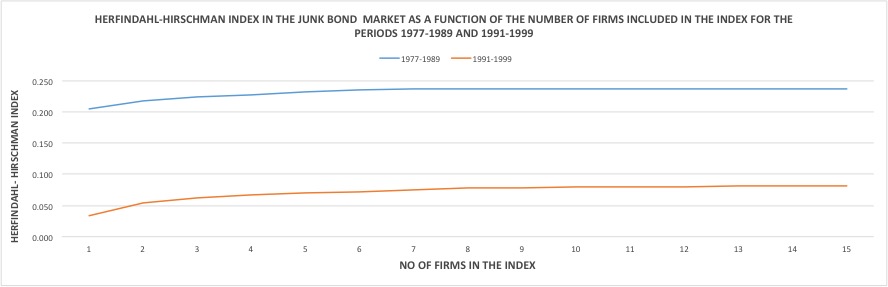}\caption{\label{fig:Hirfindahl-Hirschman-Index-for}Hirfindahl-Hirschman Index
for Junk Bonds}

\end{figure}

\begin{doublespace}
Milken helped connect entrepreneurs, growing businesses and companies
in new and unproven industries with investors. He was responsible
for the democratization of credit through the infusion of new capital.
It was widely believed around this time that companies did not need
billions in assets and strong credit ratings, all they needed was
Michael Milken. Milken identified a large market without liquidity
and provided it. There was an explosive growth in the junk bond market
and Drexel became a leading player in this market. Other firms struggled
to duplicate this success, as Milken was the focal point (Figure \ref{fig:Hirfindahl-Hirschman-Index-for};
Livingston \& Williams 2007; End-note \ref{enu:The-Herfindahl-index}).
In addition, the easy access to capital might have brought about stable
growth and economic calm to American industry in sharp contrast to
the preceding decades (Brunner 1983; Conte, Karr, Clack \& Hug 2001;
End-note \ref{enu:The-early-1980s-Recession}).

\end{doublespace}
\begin{doublespace}

\subsection{The Prestige of Michael Milken}
\end{doublespace}

\begin{doublespace}
Junk Bonds were used for LBOs and new firms emerged focused on LBOs,
creating the modern private equity (PE) industry. Milken (indirectly
through the PE firms), by eliminating corporate largess created by
conglomerates, might have generated large gains for shareholders and
the wider economy as a whole. A lot of these after effects can be
studied under the category of unintended consequences which are common
in finance and everywhere else as well (Kashyap 2016, 2016). The American
economy has been transformed since the increase of junk bonds through
the widespread use of LBOs. The following sources chronicle the spread
of this phenomenon globally and the impact it has had on the global
economic landscape: (Bruton, Dattani, Fung, Chow \& Ahlstrom 1999;
Leeds \& Sunderland 2003; Froud \& Williams 2007; Banerjee 2008; Cumming,
Fleming, Johan \& Takeuchi 2012; Lerner, Leamon \& Hardymon 2012).

An important question that arises and hints at a big involvement of
clever financial innovations is: How much of the Silicon Valley technology
boom was due to junk bonds or more subtly, due to the use of tailored
and novel means of financing? Investors witnessed that there was a
possibility for significant returns and that financial success could
be had due to the risk mitigation that Milken demonstrated by investing
in so called high risk and low profile companies. (Ooms, Werker, Caniëls,
\& Van Den Bosch 2015) suggest that regions’ development paths emerge
from agglomeration patterns and research orientation. They categorize
data from thirty-six European regions according to research orientation
and agglomeration. Clearly, they fail to identify the importance of
innovative financing as a contributing factor. (Klepper 2010) analyze
the factors behind the historical clustering of the two industries
in Detroit and Silicon Valley and consider some of the effects of
capital structure and financing. (Saxenian 2001; End-note \ref{enu:Route-128,-the})
compare Silicon Valley with the Route 128 beltway around Boston, Massachusetts
using an alternative network approach and provide important insights
into the local sources of competitive advantage (also, Hulsink, Manuel
\& Bouwman 2007). (Phillips 2001) is a study regarding law firms in
the Silicon valley region between 1946 to 1996. For other regions
in the world that are likened to Silicon valley, see: (Mathews 1997;
Patni 1999; Saxenian \& Hsu 2001; Park 2012).

\textbf{We provide additional illustrations in Appendix \ref{sec:Appendix-of-Additional}
that chronicle the growth of silicon valley and more recently other
technology centers around the world.} (Figures \ref{fig:Fastest-Growing-Electronics},
\ref{fig:Total-High-Tech}; End-note \ref{enu:Figures-SV-128}) show
the number of fastest growing electronic firms and the total employment
in technology firms in Silicon Valley and Route 128. (Figures \ref{fig:Population-Density-Silicon},
\ref{fig:Founding-Failures-Silicon}; End-note \ref{enu:Figures-SV-Law-Firms})
show the population density of law firms and the number of law firms
that were founded and failed. (Figures \ref{fig:Entry-and-Exit-IC},
\ref{fig:Percentage-of-Integrated}, \ref{fig:Entry-and-Exit-Transistor},
\ref{fig:Percentage-of-Transistor}; End-note \ref{enu:Figures-SV-IC-Transistor})
show the number of transistor and IC firms that entered and exited
the US markets; also shown are the percentage share of IC and transistor
firms across Los Angeles, Boston, New York and Silicon Valley. (Figure
\ref{fig:Growth-of-Silicon}; End-note \ref{enu:Figure-SV-Growth-1950-2000})
shows the growth of Silicon Valley in terms of the number of firms
in various constituent industries from 1950 to 2000. 

In all these figures, it is easily seen that there is increased activity
during the years when junk bonds and Micheal Milken were at their
peak. 
\end{doublespace}
\begin{doublespace}

\section{\label{sec:The-Rise-and}The Rise and Fall }
\end{doublespace}

\begin{doublespace}
Milken cherished his privacy and kept a low profile. His deals were
complex and secretive. He was an outsider in the clubby wall street
community. At this peak, he was committing more money than KKR, a
leading PE firm (Fischel 1995; End-notes \ref{enu:The-following-sources},
\ref{enu:KKR-=000026-Co.}). Using the Elbow is a foul in Soccer,
but in an LBO many people get the Elbow (they are laid off from their
jobs). Since self financing is tough, debt financing made up between
60\% to 90\% of the transactions used to acquire majority ownership
in public companies. This led to growing negative sentiment against
the junk bond industry by the American public. Some of these dynamics
are portrayed in the classic movie Wall Street, where Gordon Gekko
played by Michael Douglas is an iconic corporate raider (End-note
\ref{enu:Wall-Street-Movie}). This surge in growing negative sentiment
was despite another corner with changing attitudes which held the
belief in agency theory and the responsibility of business: Managers
should run the business to benefit stock holders (Ross 1973; Eisenhardt
1989). 
\end{doublespace}
\begin{doublespace}

\subsection{\label{subsec:Milken-the-Felon}Milken the Felon (or Not?)}
\end{doublespace}

\begin{doublespace}
With a lot of negative press and attention, the industry attracted
the attention of regulators. In the late 1980s federal authorities
investigated the business practices of one of Milken's clients, the
prominent financier and arbitrager Ivan Boesky (End-note \ref{enu:Ivan-Frederick-Boesky}).
Boesky was well-known and respected in the business world, though
he was nicknamed “Ivan the Terrible” for being a ruthless investor
(End-note \ref{enu:Ivan-Terrible}). Though as the story unfolds,
we will see that Ivan just turned out to be a terrible insider. We
all flout the connections we have in high places. But we need to realize
that people in high places are nobody’s friends; though everyone calls
them friends. This aspect of lack of cooperation among rational individuals
is studied in game theory under the heading prisoner's dilemma (Osborne
\& Rubinstein 1994; End-note \ref{enu:The-prisoner's-dilemma}). 

In 1986, Boseky was accused by federal regulators of insider trading.
Boesky paid a \$100 million penalty, and provided authorities with
information on other white collar crimes. This information led authorities
to investigate Milken in 1986 and 1987. Based on their findings, in
late 1988 federal regulators charged both Drexel and Milken in civil
court for a range of legal violations. Rudolph Giuliani, politically
ambitious, took extreme measures and invoked RICO Act (Racketeer Influenced
and Corrupt Organizations -- meant for organized crime; End-note
\ref{enu:Rudolph-William-Louis}). Drexel paid a \$650 million fine
and pleaded guilty to six felonies. Milken did not settle initially
and was indicted on 100 criminal charges. Finally he settled down
with a \$600 million fine and six charges (described by some as technical
violations, or, honest mistakes with an immaterial impact, with economic
around cost of \$318,082). 

The following questions about Milken bubble to the surface as we try
to come to terms with what he did and how much of those were actually
crimes: 
\end{doublespace}
\begin{enumerate}
\begin{doublespace}
\item Was he a greedy hustler? or Was he doing what he believed in and just
making a lot of money, perhaps, some of it through questionable means? 
\item Had he crossed the point of no return? If he stopped what he was doing,
perhaps, his entire creation could come crashing down. He was instrumental
to the eco system he had created and realizing his pivotal role, he
had to keep going.
\item He traded for the firm and many times when he traded for his own account,
was it only when the firm did not back him on certain deals? 
\item He awarded interests in his investment partnerships to portfolio managers
in return for investing in his funds. Is this a question of simple
incentives or was it much more?
\end{doublespace}
\end{enumerate}
\begin{doublespace}

\subsection{\label{subsec:Client-Loyalty,-A}Client Loyalty, A Capital Crime
(?) }
\end{doublespace}

\begin{doublespace}
Judge Kimba M. Wood’s remark when sentencing Milken to ten years in
prison was: Milken profited from his crimes, “not necessarily by lining
your pockets directly and immediately, but by increasing your clients
loyalty to you, hence, increasing, your edge over competitors and
increasing the likelihood that clients would pay for your services
in the future” (End-note \ref{enu:The-Milken-Sentence;}).

Unless the remark above was taken out of context, “increasing client
loyalty …” seems like the fundamental objective of any business. Or
perhaps, Judge Kimba Wood is referring to Utopian business entities,
where there is no competition but only co-operation (or to be more
realistic and less utopian, maybe just a combination of competition
and co-operation, termed co-opetition: Brandenburger \& Nalebuff 1996).
Perhaps such firms did exist before, and we just need to find or establish
them again.
\end{doublespace}
\begin{doublespace}

\subsection{\label{subsec:What-would-you}What would you do? What is your verdict? }
\end{doublespace}

\begin{doublespace}
We need to remember this before passing our judgment: There are no
good or bad people; just seemingly tough situations and mediocre role
models. We conclude with the following observations in Milken’s Defense: 
\end{doublespace}
\begin{enumerate}
\begin{doublespace}
\item Big companies were not competitive, so they were forced to trim (Budros
1999; Moen 1999; Gandolfi 2014).
\item Milken and his team were matching capital to entrepreneurs who could
use it effectively. As he recounted, once the dust had settled on
his legal case and he was serving his sentence, he was proud of most
of what they accomplished (Michaels \& Berman 1992).
\item After most Mergers and Acquisitions, companies find ways to increase
synergies and improve their bottomline. A primary strategy for accomplishing
this is workforce reduction (O'Shaughnessy \& Flanagan 1998; Gutknecht
\& Keys 1993; Krishnan \& Park 2002; Krishnan, Hitt \& Park 2007).
Fact: The vast majority of mergers were not financed with junk bonds.
LBOs accounted for only 15\% to 25\% (the peak 25\% happened in 1988)
of all completed mergers (Smith 1992). 
\end{doublespace}
\end{enumerate}
\begin{doublespace}

\section{\label{sec:End-notes}Acknowledgments and End-notes}
\begin{figure}[H]
\includegraphics[width=8cm]{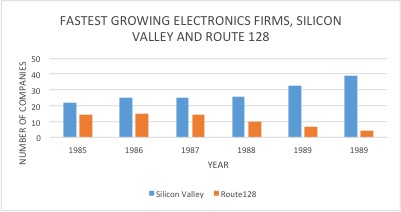}

\caption{\label{fig:Fastest-Growing-Electronics}Fastest Growing Electronics
Firms in Silicon Valley and Route 128}
\end{figure}

\begin{figure}[H]
\includegraphics[width=8cm]{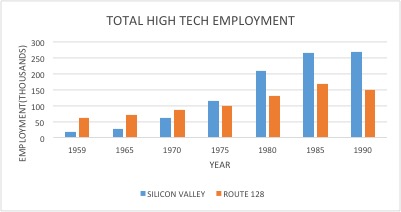}

\caption{\label{fig:Total-High-Tech}Total High Tech Employment in Silicon
Valley Firms}

\end{figure}

\begin{figure}[H]
\includegraphics[width=8cm]{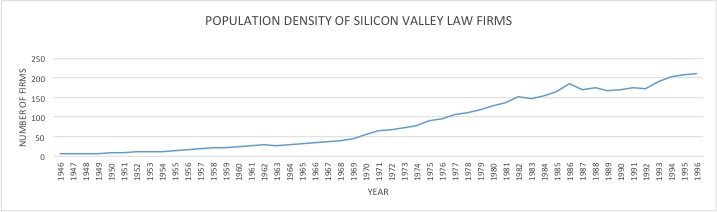}

\caption{\label{fig:Population-Density-Silicon}Population Density Silicon
Valley Law Firms}
\end{figure}

\begin{figure}[H]

\includegraphics[width=8cm]{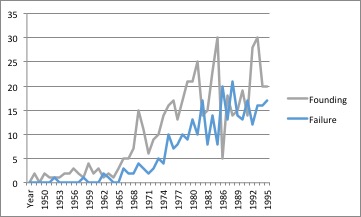}\caption{\label{fig:Founding-Failures-Silicon}Founding Failures Silicon Valley
Law Firms}

\end{figure}

\begin{figure}[H]
\includegraphics{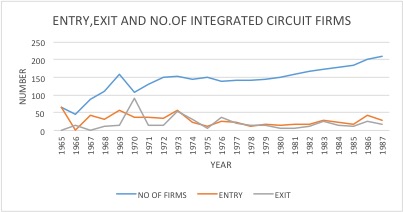}

\caption{\label{fig:Entry-and-Exit-IC}Entry and Exit of Integrated Circuit
Firms}

\end{figure}

\begin{figure}[H]
\includegraphics[width=8cm]{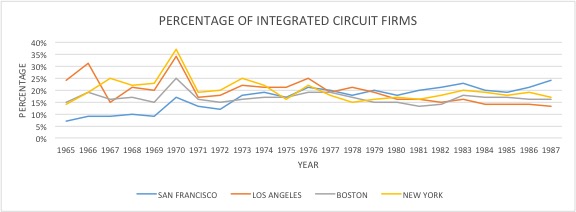}

\caption{\label{fig:Percentage-of-Integrated}Percentage of Integrated Circuit
Firms}

\end{figure}

\begin{figure}[H]
\includegraphics[width=16cm]{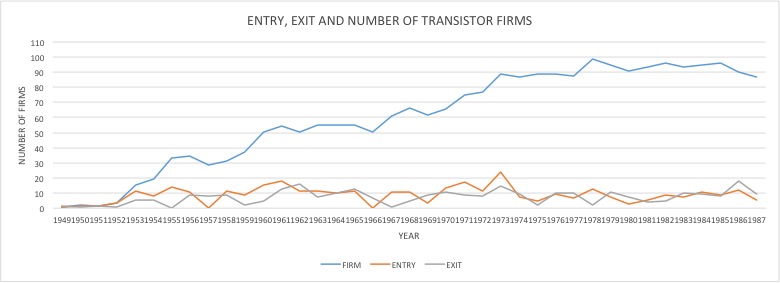}

\caption{\label{fig:Entry-and-Exit-Transistor}Entry and Exit of Transistor
Firms}

\end{figure}

\begin{figure}[H]
\includegraphics[width=16cm]{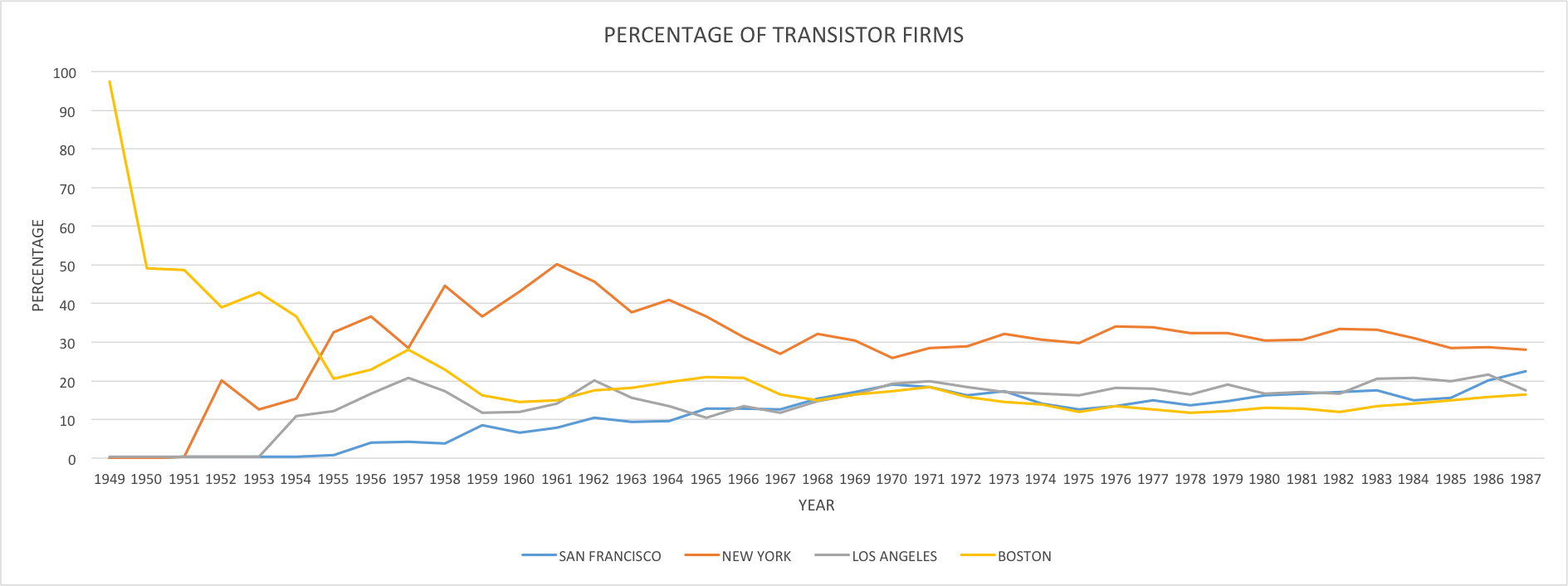}

\caption{\label{fig:Percentage-of-Transistor}Percentage of Transistor Firms}

\end{figure}

\begin{figure}[H]
\includegraphics[width=8cm]{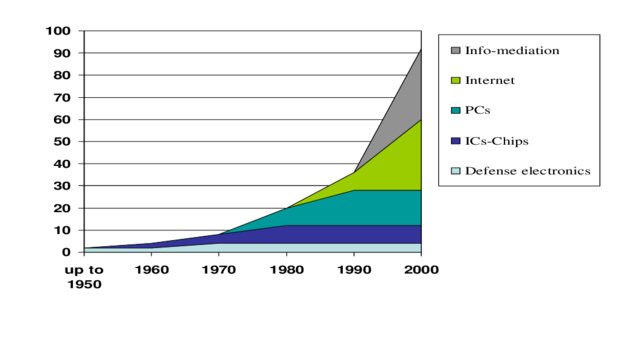}

\caption{\label{fig:Growth-of-Silicon}Growth of Silicon Valley from 1950 to
2000}

\end{figure}

\end{doublespace}
\begin{enumerate}
\begin{doublespace}
\item Naresh Pandey provided excellent assistance with gathering the numerical
illustrations used in this paper from various sources. This paper
developed over many years to supplement classroom discussions regarding
Michael Milken, Junk Bonds and Private Equity. Dr. Yong Wang, Dr.
Isabel Yan, Dr. Vikas Kakkar, Dr. Fred Kwan, Dr. William Case, Dr.
Srikant Marakani, Dr. Qiang Zhang, Dr. Costel Andonie, Dr. Jeff Hong,
Dr. Guangwu Liu, Dr. Humphrey Tung and Dr. Xu Han at the City University
of Hong Kong; the faculty members of SolBridge International School
of Business and numerous seminar participants provided encouragement
to explore and where possible apply cross disciplinary techniques.
The views and opinions expressed in this article, along with any mistakes,
are mine alone and do not necessarily reflect the official policy
or position of either of my affiliations or any other agency.
\item \label{enu:The-following-sources}The following sources have been
used to get a good understanding of Michael Milken's Background: The
Predators' Ball: the inside story of Drexel Burnham and the rise of
the junk bond raiders (Bruck 1989); Dangerous dreamers: the financial
innovators from Charles Merrill to Michael Milken (Sobel 2000); Harvard
Business School Case, Michael Milken: The Junk Bond King by Nicholas,
Tom, and Matthew G. Preble: \href{http://www.hbs.edu/faculty/Pages/item.aspx?num=50852 }{"Michael Milken: The Junk Bond King." Harvard Business School Case 816-050, March 2016.};
California State University Northridge Case, Michael Milken - The
Junk Bond King: \href{http://www.csun.edu/~hbsoc126/soc1/michael\%20milken.pdf }{California State University Northridge Case};
Wikipedia Link on Michael Milken: \href{https://en.wikipedia.org/wiki/Michael_Milken }{Michael Milken, Wikipedia Link}
\item \label{enu:Private-equity-typically}Private equity typically refers
to investment funds, generally organized as limited partnerships,
that buy and restructure companies that are not publicly traded. Private
equity is, strictly speaking, a type of equity and one of the asset
classes consisting of equity securities and debt in operating companies
that are not publicly traded on a stock exchange. However, the term
has come to be used to describe the business of taking a company into
private ownership in order to restructure it before selling it again
at a hoped-for profit. \href{https://en.wikipedia.org/wiki/Private_equity}{Private Equity, Wikipedia Link}
\item \label{enu:A-leveraged-buyout}A leveraged buyout (LBO) is a financial
transaction in which a company is purchased with a combination of
equity and debt, such that the company's cash flow is the collateral
used to secure and repay the borrowed money. \href{https://en.wikipedia.org/wiki/Leveraged_buyout}{Leveraged Buyout, Wikipedia Link}
\item \label{enu:Figure-Market-Junk}Figure \ref{fig:The-Market-for-Junk}
was prepared using data from (Altunbaş, Gadanecz \& Kara 2006a).
\item \label{enu:Figures-Drexel-Fees}Figures \ref{fig:Spread-The-Junk},
\ref{fig:Drexel-Underwriting-Fees}, \ref{fig:The-Market-for-Junk},
\ref{fig:Hirfindahl-Hirschman-Index-for} were prepared using data
from (Livingston \& Williams 2007).
\item \label{enu:Figures-SV-128}Figures \ref{fig:Fastest-Growing-Electronics},
\ref{fig:Total-High-Tech} were prepared using data from (Saxenian
2001). The fastest growing companies chart (Figure \ref{fig:Fastest-Growing-Electronics})
is based on the 100 fastest-growing electronics firms in the United
States of America.
\item \label{enu:Figures-SV-Law-Firms}Figures \ref{fig:Population-Density-Silicon},
\ref{fig:Founding-Failures-Silicon} were prepared using data from
(Phillips 2001).
\item \label{enu:Figures-SV-IC-Transistor}Figures \ref{fig:Entry-and-Exit-IC},
\ref{fig:Percentage-of-Integrated}, \ref{fig:Entry-and-Exit-Transistor},
\ref{fig:Percentage-of-Transistor} were prepared using data from
(Klepper 2010).
\item \label{enu:Figure-SV-Growth-1950-2000}Figure \ref{fig:Growth-of-Silicon}
was prepared using data from (Hulsink, Manuel \& Bouwman 2007).
\item \label{enu:=00201CFinancing-is-an}“Financing is an art form. One
of the challenges is how to correctly finance a company. In certain
periods of time, more covenants need to be put into deals. You have
to be sure the company has the right covenant -{}- to allow it the
freedom to grow, but also to insure the integrity of the credit. Sometimes
a company should issue convertible bonds instead of straight bonds.
Sometimes it should issue preferred stock. Each company and each financing
is different, and the process can’t be imitative.”\textbf{ }\href{https://www.goodreads.com/author/quotes/55100.Michael_Milken }{Good Reads Quote}
\item \label{enu:Ghostface-Killah-=000026}Milk It: To extract the most
out of a situation. An abbreviated version of “milk it for all it’s
worth” \href{https://www.urbandictionary.com/define.php?term=milk\%20it}{Milk It, Urban Dictionary};
We adapt milk it to our situation and write the following, ``Michael
Milken was Milking Them (Milk'em) Junk Bonds''; Also there is a famous
song somewhat related to this theme called Milk'em: \href{https://www.youtube.com/watch?v=wz-x9FBDmyY}{Ghostface Killah \& Trife Da God - Milk ’em Song};
\href{https://genius.com/Ghostface-killah-and-trife-da-god-milk-em-lyrics }{Milk’em Lyrics}
\item \label{enu:Again-School-Princess}Again, here we cannot be sure what
this really means since we are unsure when this was said and by whom.
For one, who is not a school princess. Also, we need to keep in mind
that people are known to say nice things to a lady whose husband is
earning 500 million dollars a year.
\item \label{enu:Voted-most-likely}While there is still a lot of ambiguity,
voted most likely to succeed is a recorded fact and does indicate
that the person was likely to be ambitious and capable.
\item \label{enu:California-Songs}The following is a list of Songs about
California: \href{https://en.wikipedia.org/wiki/List_of_songs_about_California}{Songs About California}
\item \label{enu:Again-Credit-Compensation}Again, we need to be mindful
that many would say nice things to someone who is instrumental in
ensuring that they earn very high compensations.
\item \label{enu:Tipping-Point}A tipping point is defined as \textquotedbl the
moment of critical mass, the threshold, the boiling point\textquotedbl .
\href{https://en.wikipedia.org/wiki/The_Tipping_Point}{The Tipping Point, Wikipedia Link}
\item \label{enu:The-Prestige-is}The Prestige is a 2006 psychological thriller
film directed adapted from a novel of the same name; \href{https://en.wikipedia.org/wiki/The_Prestige_(film)}{The Prestige (Film), Wikipedia Link};
The Prestige on IMDb: \href{http://www.imdb.com/title/tt0482571/quotes}{Prestige on IMDB} 
\item \label{enu:The-Herfindahl-index}The Herfindahl index (also known
as Herfindahl--Hirschman Index, HHI, or sometimes HHI-score) is a
measure of the size of firms in relation to the industry and an indicator
of the amount of competition among them. It is defined as the sum
of the squares of the market shares of the firms within the industry
(sometimes limited to a certain number of the largest firms), where
the market shares are expressed as fractions. The result is proportional
to the average market share, weighted by market share. As such, it
can range from 0 to 1.0, moving from a huge number of very small firms
to a single monopolistic producer. Increases in the Herfindahl index
generally indicate a decrease in competition and an increase of market
power, whereas decreases indicate the opposite. \href{https://en.wikipedia.org/wiki/Herfindahl–Hirschman_Index}{Herfindahl–Hirschman Index, Wikipedia Link}
\item \label{enu:The-early-1980s-Recession}The early 1980s recession was
a severe global economic recession that affected much of the developed
world in the late 1970s and early 1980s. Clearly, numerous factors
contribute to complex events such as recessions. Following are some
sources that summarize the main events around this time: (Day 1993;
Murray 2008); \href{https://en.wikipedia.org/wiki/Early_1980s_recession}{Early 1980s Recession, Wikipedia Link}
\item \label{enu:Route-128,-the}Route 128, the Yankee Division Highway,
is a state highway in the U.S. state of Massachusetts. The area along
the western part of Route 128 is home to a number of high-technology
firms and corporations. This part of Route 128 was dubbed \textquotedbl America's
Technology Highway\textquotedbl . In the 1980s, the area was often
compared to California's Silicon valley. \href{https://en.wikipedia.org/wiki/Massachusetts_Route_128}{Massachusetts Route 128, Wikipedia Link}
\item \label{enu:KKR-=000026-Co.}KKR \& Co. Inc. (formerly known as Kohlberg
Kravis Roberts \& Co. and KKR \& Co. L.P.) is a global investment
firm that manages multiple alternative asset classes, including private
equity, energy, infrastructure, real estate, credit, and, through
its strategic partners, hedge funds. \href{https://en.wikipedia.org/wiki/Kohlberg_Kravis_Roberts}{Kohlberg Kravis Roberts, Wikipedia Link}
\item \label{enu:Wall-Street-Movie}Wall Street is a 1987 American drama
film, directed and co-written by Oliver Stone, which stars Michael
Douglas, Charlie Sheen, and Daryl Hannah. The film tells the story
of Bud Fox (Sheen), a young stockbroker who becomes involved with
Gordon Gekko (Douglas), a wealthy, unscrupulous corporate raider.
\href{https://en.wikipedia.org/wiki/Wall_Street_(1987_film)}{Wall Street (1987 Film), Wikipedia Link}
\item \label{enu:Ivan-Frederick-Boesky}Ivan Frederick Boesky (born March
6, 1937) is a former American stock trader who became infamous for
his prominent role in an insider trading scandal that occurred in
the United States during the mid-1980s. \href{https://en.wikipedia.org/wiki/Ivan_Boesky}{Ivan Boesky, Wikipedia Link}
\item \label{enu:Ivan-Terrible}Ivan IV Vasilyevich (25 August 1530 --
28 March 1584), commonly known as Ivan the Terrible (\textquotedbl Ivan
the Formidable\textquotedbl{} or \textquotedbl Ivan the Fearsome\textquotedbl ),
was the Grand Prince of Moscow from 1533 to 1547 and the first Tsar
of Russia from 1547 to 1584. \href{https://en.wikipedia.org/wiki/Ivan_the_Terrible}{Ivan The Terrible, Wikipedia Link}
\item \label{enu:The-prisoner's-dilemma}The prisoner's dilemma is a standard
example of a game analyzed in game theory that shows why two completely
rational individuals might not cooperate, even if it appears that
it is in their best interests to do so. \href{https://en.wikipedia.org/wiki/Prisoner's_dilemma}{Prisoner's Dilemma, Wikipedia Link}
\item \label{enu:Rudolph-William-Louis}Rudolph William Louis Giuliani (born
May 28, 1944) is an American politician, attorney, businessman, and
public speaker who served as the 107th Mayor of New York City from
1994 to 2001. \href{https://en.wikipedia.org/wiki/Rudy_Giuliani}{Rudy Giuliani, Wikipedia Link}
\item \label{enu:The-Milken-Sentence;}The Milken Sentence; Excerpts From
Judge Wood's Explanation of the Milken Sentencing: \href{https://www.nytimes.com/1990/11/22/business/milken-sentence-excerpts-judge-wood-s-explanation-milken-sentencing.html}{NY Times Article, Nov 22, 1990}
\end{doublespace}
\end{enumerate}
\begin{doublespace}

\section{\label{sec:References}References}
\end{doublespace}
\begin{enumerate}
\begin{doublespace}
\item Altunbaş, Y., Gadanecz, B., \& Kara, A. (2006a). Banks’ and Financial
Institutions’ Decision to Participate in Loan Syndications. In Syndicated
Loans (pp. 101-125). Palgrave Macmillan, London.
\item Altunbaş, Y., Gadanecz, B., \& Kara, A. (2006b). The evolution of
syndicated loan markets. The Service Industries Journal, 26(6), 689-707.
\item Banerjee, A. (2008). Private equity in developing nations. Journal
of Asset Management, 9(2), 158-170.
\end{doublespace}
\item Brandenburger, A. M., \& Nalebuff, B. J. (1996). Co-opetition. Crown
Business.
\begin{doublespace}
\item Bruck, C. (1989). The Predators' Ball: the inside story of Drexel
Burnham and the rise of the junk bond raiders. Penguin Group USA.
\item Budros, A. (1999). A conceptual framework for analyzing why organizations
downsize. Organization Science, 10(1), 69-82.
\item Brunner, K. (1983). The Recession of 1981/1982 in the Context of Postwar
Recessions.
\item Bruton, G. D., Dattani, M., Fung, M., Chow, C., \& Ahlstrom, D. (1999).
Private equity in China: Differences and similarities with the Western
model. The Journal of Private Equity, 2(2), 7-13.
\item Conte, C., Karr, A. R., Clack, G., \& Hug, K. E. (2001). Outline of
the US Economy. US Department of State, Office of International Information
Programs.
\item Cumming, D., Fleming, G., Johan, S., \& Takeuchi, M. (2012). Legal
protection, corruption and private equity returns in Asia. In Entrepreneurship,
Governance and Ethics (pp. 173-193). Springer, Dordrecht.
\item Day, K. (1993). S \& L Hell: The people and the politics behind the
\$1 trillion savings and loan scandal. WW Norton \& Company.
\item Eisenhardt, K. M. (1989). Agency theory: An assessment and review.
Academy of management review, 14(1), 57-74.
\item Fischel, D. R. (1995). Payback: the conspiracy to destroy Michael
Milken and his financial revolution. New York: HarperBusiness.
\item Froud, J., \& Williams, K. (2007). Private equity and the culture
of value extraction. New Political Economy, 12(3), 405-420.
\item Gandolfi, F. (2014). Why do firms downsize?. Journal of Management
Research (09725814), 14(1).
\item Gutknecht, J. E., \& Keys, J. B. (1993). Mergers, acquisitions and
takeovers: Maintaining morale of survivors and protecting employees.
Academy of Management Perspectives, 7(3), 26-36.
\item Hickman, W. B. (1958). Corporate bond quality and investor experience.
NBER Books.
\item Hulsink, W., Manuel, D., \& Bouwman, H. (2007). Clustering in ICT:
From Route 128 to Silicon Valley, from DEC to Google, from hardware
to content. ERIM Report Series Reference No. ERS-2007-064-ORG.
\item Kaplan, S. N., \& Stromberg, P. (2009). Leveraged buyouts and private
equity. Journal of Economic Perspectives, 23(1), 121-46.
\item Kashyap, R. (2015). A Tale of Two Consequences. The Journal of Trading,
10(4), 51-95. 
\item Kashyap, R. (2016). Hong Kong--Shanghai Connect/Hong Kong--Beijing
Disconnect? Scaling the Great Wall of Chinese Securities Trading Costs.
The Journal of Trading, 11(3), 81-134.
\item Kashyap, R. (2019). For Whom the Bell (Curve) Tolls: A to F, Trade
Your Grade Based on the Net Present Value of Friendships with Financial
Incentives. The Journal of Private Equity, 22(3), 64-81.
\item Klepper, S. (2010). The origin and growth of industry clusters: The
making of Silicon Valley and Detroit. Journal of Urban Economics,
67(1), 15-32.
\item Krishnan, H. A., \& Park, D. (2002). The impact of work force reduction
on subsequent performance in major mergers and acquisitions: an exploratory
study. Journal of Business Research, 55(4), 285-292.
\item Krishnan, H. A., Hitt, M. A., \& Park, D. (2007). Acquisition premiums,
subsequent workforce reductions and post‐acquisition performance.
Journal of Management Studies, 44(5), 709-732.
\item Leeds, R., \& Sunderland, J. (2003). Private equity investing in emerging
markets. Journal of applied corporate finance, 15(4), 111-119.
\item Lerner, J., Leamon, A., \& Hardymon, F. (2012). Venture capital, private
equity, and the financing of entrepreneurship.
\item Livingston, M., \& Williams, G. (2007). Drexel Burnham Lambert's bankruptcy
and the subsequent decline in underwriter fees. Journal of Financial
Economics, 84(2), 472-501.
\item Mathews, J. A. (1997). A Silicon Valley of the East: Creating Taiwan's
semiconductor industry. California Management Review, 39(4), 26-54.
\item Michaels, J. W., \& Berman, P. (1992). My story-{}-Michael Milken.
Forbes, 149(6), 78-92.
\item Murray, C. (2008). Losing ground: American social policy, 1950-1980.
Basic books.
\item Moen, Ø. (1999). The relationship between firm size, competitive advantages
and export performance revisited. International Small Business Journal,
18(1), 53-72.
\item Ooms, W., Werker, C., Caniëls, M. C., \& Van Den Bosch, H. (2015).
Research orientation and agglomeration: Can every region become a
Silicon Valley?. Technovation, 45, 78-92.
\item Osborne, M. J., \& Rubinstein, A. (1994). A course in game theory.
MIT press.
\item O'Shaughnessy, K. C., \& Flanagan, D. J. (1998). Determinants of layoff
announcements following M\&As: An empirical investigation. Strategic
management journal, 19(10), 989-999.
\item Park, S. C. (2012). Competitiveness of East Asian science cities:
discourse on their status as global or local innovative clusters.
Ai \& Society, 27(4), 451-464.
\item Patni, A. (1999). Silicon valley of the east. Harvard International
Review, 21(4), 8.
\item Phillips, D. J. (2001). The promotion paradox: Organizational mortality
and employee promotion chances in Silicon Valley law firms, 1946--1996.
American Journal of Sociology, 106(4), 1058-1098.
\item Ross, S. A. (1973). The economic theory of agency: The principal's
problem. The American economic review, 63(2), 134-139.
\item Ross, S. A., Westerfield, R., \& Jaffe, J. F. (2013). Corporate finance.
McGraw-Hill/Irwin.
\item Saxenian, A. (2001). Inside-out: regional networks and industrial
adaptation in Silicon Valley and Route 128. The sociology of economic
life, 2, 357-375.
\item Saxenian, A., \& Hsu, J. Y. (2001). The Silicon Valley--Hsinchu connection:
technical communities and industrial upgrading. Industrial and corporate
change, 10(4), 893-920.
\item Smith, R. C. (1992). After the Ball. The Wilson Quarterly (1976-),
16(4), 31-43.
\end{doublespace}
\item Sobel, R. (2000). Dangerous dreamers: the financial innovators from
Charles Merrill to Michael Milken. Beard Books.
\begin{doublespace}
\item Gladwell, M. (2006). The tipping point: How little things can make
a big difference. Little, Brown.
\item White, L. J. (2010). Markets: The credit rating agencies. Journal
of Economic Perspectives, 24(2), 211-26.
\end{doublespace}
\end{enumerate}
\begin{doublespace}

\section{\label{sec:Appendix-of-Additional}Appendix of Additional Illustrations}
\end{doublespace}

\begin{doublespace}


\end{doublespace}

\end{document}